\documentclass[%
 reprint,
superscriptaddress,
 amsmath,amssymb,
 aps,
]{revtex4-1}

\usepackage{graphicx}
\usepackage{dcolumn}
\usepackage{bm}

\usepackage{hyperref}

\usepackage{commath}
\usepackage{xcolor}
\usepackage{tabu}

\graphicspath{ {./} }
\DeclareGraphicsExtensions{.pdf}

\usepackage{subfig}
\begin{document}


\title{Topological defects of dipole patchy particles on a spherical surface}

\author{Uyen Tu Lieu}
\email{uyen.lieu@aist.go.jp}
\affiliation{Mathematics for Advanced Materials-OIL, AIST, 2-1-1 Katahira, Aoba, 980-8577 Sendai, Japan}

\author{Natsuhiko Yoshinaga}
\email{yoshinaga@tohoku.edu.jp} 
\affiliation{Mathematics for Advanced Materials-OIL, AIST, 2-1-1 Katahira, Aoba, 980-8577 Sendai, Japan}
\affiliation{WPI-Advanced Institute for Materials Research (WPI-AIMR), Tohoku University, 2-1-1 Katahira, Aoba, 980-8577 Sendai, Japan}

\date{\today}

\begin{abstract}
We investigate the assembly of the dipole-like patchy particles confined to a spherical surface by Brownian dynamics simulations. The surface property of the spherical particle is described by the spherical harmonic $Y_{10}$, and the orientation of the particle is defined as the uniaxial axis. On a flat space, we observe a defect-free square lattice with nematic order. On a spherical surface, defects appear due to the topological constraint. As for the director field, four defects of winding number $+1/2$ are observed, satisfying the Euler characteristic. We have found many configurations of the four defects lying near a great circle. Regarding the positional order for the square lattice, eight grain boundary scars proliferate linearly with the sphere size. The positions and orientations of the eight grain boundary scars are strongly related to the four $+1/2$ defect cores.
\end{abstract}

\maketitle



\section{\label{sec:Introduction}Introduction}
Patchy particles are particles of colloidal size and have patches playing as specific interactive sites on the particles. Due to the anisotropic interaction, patchy particles are capable of assembling into complex structures whose properties are fundamentally different from the ``conventional'' materials \cite{chen_directed_2011}. Recent developments in synthesis techniques \cite{chen_triblock_2011,wang_colloids_2012,choueiri_surface_2016} have made it feasible to fabricate patchy particles of high degrees of freedom \cite{chen_triblock_2011,romano_phase_2010,romano_phase_2012,romano_patterning_2012,hong_clusters_2006,hong_clusters_2008} and attracted research on the self-assembly of patchy particles. Practical applications of the particle assembly require knowledge of the principal design of the particle for a specific structure and vice versa, and how to produce high yield target structures. Understanding the principles of control defect is important for those tasks. Defects are imperfections and singularities in an ordered structure. From a microscopic point of view, defects can be served as interacting sites for chemical linkers to promote the formation of large scale structures \cite{nelson_toward_2002, zhang_self-assembly_2005}. On the macroscopic length-scale, defects are inevitable during crystallisation and affect the overall properties. In particular, using curvature is one method to study defects because the defects are topologically protected, meaning that they can not disappear by continuous deformation of the order parameter. We focus on spherical surface for its substantial applicability such as crystalline membrane, design of crystalline materials, fabrication of patchy colloids. 

The in-plane order of a two-dimensional crystal on a curved surface is a much more complicated problem than that on a flat space because the geometrical and topological constraints have to be taken into account. The ground state of isotropic particles on two-dimensional plane is a defect-free triangular lattice containing 6-fold coordinated particles. On a spherical surface, such a translational symmetry is broken, at least twelve 5-fold coordinated particles are required to compensate the topology of the sphere. This is similar to the truncated icosahedron pattern of a soccer ball whose twelve pentagons are icosahedrally arranged among the hexagons \cite{bowick_two-dimensional_2009, guerra_freezing_2018}. Aside from affecting the positional order of the particle system, the curved surface also influences the orientational order. For example, the rod-like particles confined to a spherical surface form four $+1/2$ or two $+1$ defects \cite{lubensky_orientational_1992,nelson_toward_2002, lopez-leon_frustrated_2011} instead of a defect-free nematic phase (long-range orientational but no long-range positional order on a flat space). 

Patchy particles have both positional and orientational order, which has been studied only separately. Although the appearance of the topological defects is generally accepted as a consequence of the embedded geometry and topology, it remains unclear about: the underlying mechanism on the formation of defects and the dependence on the system size, the influence of the particle-particle and particle-curvature interaction to the defect, and the interplay between positional and orientational order. In order to address the above problems, we consider the most simple form of anisotropic particle exhibiting both positional and orientational order, that is a spherical particle with dipole-like patches and behaves somewhat similar to magnetic bead. We dynamically simulate and compare the assembly of such patchy particles confined to a planar geometry and a spherical surface.     

\textbf{Topological defects on curved surface}\\
It is useful to discuss about the basics and the relevant studies on the topological defects of ordered structures embedded in curved surfaces. Suppose a closed surface is facetted and divided into a number of $V$ vertices, $E$ edges, $F$ faces, Euler theorem states that $V-E+F=\chi$ where the Euler characteristic $\chi=2(1-g)$, and $g$ is the genus of the closed surface. For instance, $V-E+F=2$ is applied to all polyhedra because they are topologically equal to a sphere with $\chi=2$ \cite{hyde_language_1997}. If we restrict every face has $c$ vertices, and let $N_z$ be the number of vertices which have $z$ connections with the others, then the Euler theorem can be written as (see Appendix \ref{sec:Appendix_Euler})
	\begin{equation}
	\sum_{z}{ \left( \frac{2c}{c-2} -z \right)N_z } =\frac{2c}{c-2} \chi  \label{eq:Euler}
	\end{equation}
Consider point particles on a surface triangularly facetted ($c=3$) by the points, it is well known that on a flat space the isotropic particles, most of the time, form a triangular lattice hence the 6-fold symmetry with $z=6$. A disclination in this case refers to a vertex whose $z$ deviates from six and the charge of the disclination is defined as $6-z$ \cite{nelson_defects_2002}. When such a particle system is confined to a spherical surface, it is straightforward from eqn (\ref{eq:Euler}) that a net charge of $\sum_{z}{ \left( 6-z \right)N_z } =12$ is required. For sphere size below a critical value, the twelve 5-fold coordinated particles are icosahedrally located among the otherwise 6-fold ones owning to the repulsion of the like-sign charge \cite{nelson_defects_2002}. As the sphere increases, those 5-fold disclinations are screened by additional dislocations which are pairs of 5- and 7-fold coordinated particles \cite{bowick_interacting_2000}, noted that it is not limit to the 4- or 8-fold particles as long as the sum of charge is neutral. The interaction between the clouds of dislocations eventually leads to the formation of twelve grain boundary scars where each scar consists of pairs of 5- and 7-fold coordinated particles with a net charge of $+1$. The grain boundary scars in spherical crystal are observed in both experiments and simulation \cite{einert_grain_2005,bowick_two-dimensional_2009,guerra_freezing_2018}. The interaction between the disclinations is believed to be screened and less important. The arrangement of these twelve defect scars is unusual, complex and not fully understood especially in the large sphere limit \cite{bowick_interacting_2000}. For a square lattice on the sphere, the evidence of how disclinations distribute is still lacking. However, it can be conjectured that by employing a quadrilateral mesh ($c=4$) for square lattice, a disclination is now the vertex whose connection differs from four. Eqn (\ref{eq:Euler}) becomes $\sum_{z}{ \left( 4-z \right)N_z }=8$, meaning that at least eight disclinations of $z=3$ is required for a square lattice on a sphere. The development of dislocations which are now viewed as bound disclination pairs of 3- and 5-fold coordinated particles \cite{strandburg_two-dimensional_1988} are expected to occur as the system size increases. 

The in-plane orientation on a two-dimensional surface is described by a unit vector field tangent to the surface. In general, $p$-atic vectors are invariant under rotations of $2n\pi/p$ ($n$ is integer) about the surface normal \cite{lubensky_orientational_1992}. The strength of a defect in this case is characterised via winding number $k$, defined as how much rotation of the vector field around a counter-clockwise circuit enclosing the defect core on the order parameter space, $k=\Delta\theta/(2\pi/p)$ where $\Delta\theta$ is the angle of the vector rotates in one counter-clockwise circuit \cite{chaikin_principles_1995}. On a closed surface, the net strength is equal to Euler characteristic according to Poincar\'{e}-Hopf theorem \cite{kamien_geometry_2002}. For instance, the in-plane order structure of a vector field ($p=1$) on a sphere requires at least two $+1$ defects at the two opposite poles, nematics ($p=2$) have either four $+1/2$ or two $+1$ defects. The detailed configuration of spherical $p$-atic order depends on the interaction energy of the system. According to Frank free energy approach for spherical nematics, in the one elastic constant limit, \textit{i.e.} splay constant equals bending constant, the ground state exposes four $+1/2$ defects at the vertices of a tetrahedron inscribed in the sphere \cite{lubensky_orientational_1992, nelson_toward_2002,vitelli_nematic_2006}. In contrast, in the extreme limit when splay is much softer (or harder) than bending, it is suggest that a $+1$ defect can split into two $+1/2$ defect without costing energy. As a result, an infinite number of states of four $+1/2$ defects lying near a great circle can be obtained by the cut-and-rotate surgery of the sphere with two $+1$ defects \cite{shin_topological_2008}. Another relevant case is spherical tetratics ($p=4$) where the particles are square-shaped or cross-shaped. The low energy states consist of eight $+1/4$ defects which may position on the vertices of an anti-cube \cite{lubensky_orientational_1992, li_topological_2013} or cube \cite{manyuhina_forming_2015,wang_interplay_2018}. Such difference is perhaps caused by the different types of interaction for simulation and the forms of free energy for continuous description. It is worth noting that the $p$-atic particles in those mentioned systems has purely rotational degree of freedom, hence the existence of positional order remains unclear.     

\section{\label{sec:Methods}Methods} 
\subsection{\label{sec:BD}Brownian Dynamics} 
We employ the Brownian dynamics simulation algorithm for particles in Euclidean space in overdamped limit \cite{ermak_brownian_1978, dickinson_brownian_1985}. The model is successfully applied for particles confined to a flat space. The translational and rotational motions of the particles confined to a spherical surface are given as follows 
	\begin{subequations} 
	\begin{align}
	\bm{r}(t+\Delta t) 			&=\bm{r}(t)+\frac{D^{T}}{k_B T}\bm{F}(t)\Delta t+\delta \bm{r} + \bm{F}^H 	\label{eq:Brownian_T}\\
	\bm{\Omega}(t+\Delta t) 	&=\bm{\Omega}(t)+\frac{D^{R}}{k_B T}\bm{T}(t)\Delta t+\delta \bm{\Omega} 	 \label{eq:Brownian_R}
	\end{align}
	\end{subequations}
where $\bm{r}(t+\Delta t)$, $\bm{\Omega}(t+\Delta t)$ denote the position and orientation of the particle after the time step $\Delta t$; $D^{T}$, $D^{R}$  are the translational and rotational diffusion coefficients of an isolated particle, respectively; $\delta \bm{r}$, $\delta \bm{\Omega}$ are the translation and rotation due to thermal fluctuation, satisfying $\delta \bm{r}=\bm{\delta}^G \sqrt{2D^{T}\Delta t}$, $\delta \bm{\Omega}=\bm{\delta}^G\sqrt{2D^{R}\Delta t}$ where each component of $\delta^G_i$ is independently chosen from a Gaussian distribution with zero mean and unit variance. The force $\bm{F}$ and torque $\bm{T}$ are derived from the pairwise potential. In order to capture the dynamics of particle position on the tangent plane, we apply the algorithm in Ref \cite{castro-villarreal_brownian_2014}, in which the tangential parts of the interacting force $\bm{F}$ and noise $\delta\bm{r}$ in eqn (\ref{eq:Brownian_T}) at the point where particle is located are considered; finally an addition harmonic terms $\bm{F}^H=\kappa(r-R)\bm{r}/r$ are added to enforce the confinement after the translation of each time step.  

\begin{figure}[ht] 
   \centering
   \includegraphics[width=0.45\textwidth,trim=20mm 125mm 80mm 20mm,clip]{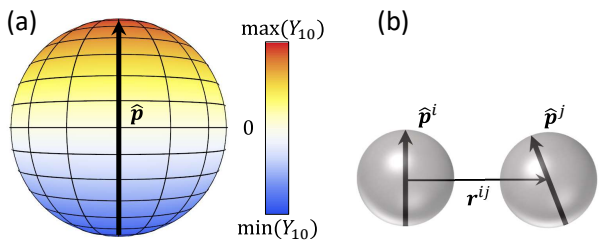}
   \caption{Illustration of (a) patterned surface of the $Y_{10}$ particle, (b) a pair of particles characterised by their orientations and relative position.}
   \label{fig:1}
\end{figure}

\begin{figure}[ht] 
   \centering
   \includegraphics[width=0.45\textwidth,trim=25mm 28mm 40mm 30mm,clip]{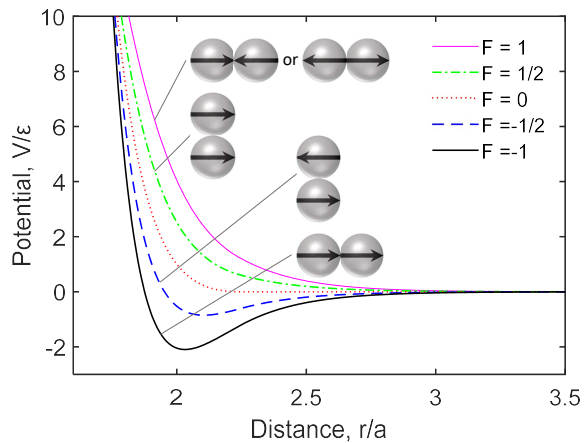}
   \caption{Pairwise potential as a function of the centre distance for some configurations.}
   \label{fig:2}
\end{figure}

The patchy particle possesses patterned surface related to its physical or chemical properties, which induces the anisotropic interaction of the particles. Such a pattern can be systematically described by means of spherical harmonics $Y_{lm}$. In this study, the dipole-like pattern of a unit particle of radius $a=1$ is given as $Y_{10}(\bm{\hat{x}})=\sqrt{3/(4\pi)}\bm {\hat{p}}\cdot \bm{\hat{x}}$ where $\bm{\hat{p}}$ is defined as the orientation of the particle (Fig. \ref{fig:1}). This pattern has positive and negative hemispheres similar to the Janus particle \cite{hong_clusters_2006,moghani_self-assembly_2013,delacruz-araujo_rich_2016}. The interaction potential for a pair of particles $i$ and $j$ comprises an isotropic Week-Chandler-Anderson potential $V_{WCA}$ preventing the overlapping of particle, and an orientation-dependent Morse potential $V_M$: 
	\begin{equation}
	V=V_{WCA}(r) - F(\bm{\hat{p}}^i,\bm{\hat{p}}^j,\bm{\hat{r}})V_M(r) 
	\end{equation}
where $\bm{r}^{ij}=\bm{r}^j-\bm{r}^i$ is the distance vector between particle centre, $r=\left|\bm{r}^{ij}\right|$, and $\bm{\hat{r}}=\bm{r}^{ij}/r$, the unit vector $\bm{\hat{p}}^i$, $\bm{\hat{p}}^j$ is the director of particles $i$, $j$, respectively, and
	\begin{equation}
	V_{WCA}= 
  	\begin{cases}
  	4\varepsilon \left[ (\frac{2a}{r})^{12}- (\frac{2a}{r})^{6}+\frac{1}{4} \right]	, & r\leq 2a\sqrt[6]{2} \\
  	0 												    							, & r > 2a\sqrt[6]{2}
  	\end{cases} 
	\end{equation}
				\begin{equation}
				V_M= \varepsilon M_d \left \{ \left[ 1-\exp{\left( -\frac{r-r_{eq}}{M_r} \right)} \right]^2 -1 \right \} 
				\end{equation}
where $\varepsilon$ is the potential well depth, $M_d$ is the Morse potential depth factor ($M_d=2.294a$), $M_r$ is the Morse potential range parameter $M_r=a$, and $r_{eq}$ is the Morse potential equilibrium position ($r_{eq}=1.878a$) \cite{delacruz-araujo_rich_2016}. The anisotropic function $F(\bm{\hat{p}}^i,\bm{\hat{p}}^j,\bm{\hat{r}})$ depends only on the mutual orientation of the particles and is given as   
\begin{equation} 
	F=-\frac{3}{2}[(\bm{\hat{p}}^i.\bm{\hat{r}}) (\bm{\hat{p}}^j.\bm{\hat{r}}) - \frac{1}{3}\bm{\hat{p}}^i.\bm{\hat{p}}^j)]
\end{equation}
$F$ is normalised so that $-1\leq F \leq 1$, the negative/positive $F$ indicates attractive/repulsive interaction. The illustration of the pair potential for some given configurations is given in Fig. \ref{fig:2}. The most favourable pair is correspondent to the head to tail arrangement. When the head to tail positions are occupied, the secondary stable structure appears in the form of side-by-side anti-parallel alignment.   

The simulation is conducted in the dimensionless form, where the characteristic length, energy, time is $a$, $\varepsilon$, $a^2/D^T$, respectively. The time step is taken so that under the condition of one unit of force and $k_BT/\varepsilon=0.1$ the particle on average moves $10^{-3}a$ in one step. The packing fraction $\rho$ is defined as the ratio of the volume of particles to that of the space confining them, which is $L \times L\times 2a$ for the planar geometry and $4/3\pi[(R+a)^3-(R-a)^3]$  for the sphere of radius $R$. The periodic boundary condition in $L$ direction is applied for the planar geometry case. The initial positions and orientations of the particles are randomly distributed. The temperature $k_BT/\varepsilon$ decreases from 0.5 to 0.05 by intervals of 0.01, and $0.5\times 10^6$ simulation steps is performed for each value of $k_BT/\varepsilon$. In this article we show the structures at the last time step. Verification of the simulation on repulsive isotropic particles is performed by setting the anisotropic function fixed as $F=1$. 

Last but not least, the particles are possibly fluctuates between certain thickness of the spherical layer. This effect, however, is negligible thanks to the small enough time step and suitable harmonic potential which is comparable to the particle-particle interaction. The systems in the study contain many interacting particles, there is possibility that metastable states are obtained instead of true ground states. We report a statistical result including all possible states.

\subsection{\label{sec:Analysis}Structure analysis} 
The coordination number $N_i$ of particle $i$ is estimated by combining the conventional Delauney triangulation whose vertices are the particle positions \cite{guerra_freezing_2018}, and a distance constraint. The particles further than a certain value are not considered as neighbours due to the relatively short-range interaction (Fig. \ref{fig:2}). Such distance constraint is based on the first trough of the pairwise distance distribution, which is approximately $2.5a$. Then the local positional order of particle $i$ is calculated by the two-dimensional bond-orientational order parameter $\psi_n$ \cite{nelson_defects_2002,guerra_freezing_2018}: 
	\begin{equation}
	\psi_n (i)=\frac{1}{N_i} \sum_{j=1}^{N_i} e^{in\theta_{ij}} 
	\end{equation}
where $\theta_{ij}$ is the angle between particle $i$ and its neighbouring particle $j$ on the tangent plane of particle $i$ if the particles are confined to the spherical surface. $|\psi_n|$ characterises the local degree of the regular $n$-gon order around a particle; for instance, the perfect square lattice on flat space has $|\psi_4|=1$, while the hexagonal one has $|\psi_6|=1$. 

\begin{figure}[ht]
    \centering
    \includegraphics[width=0.46\textwidth,trim=35mm 107mm 20mm 50mm,clip]{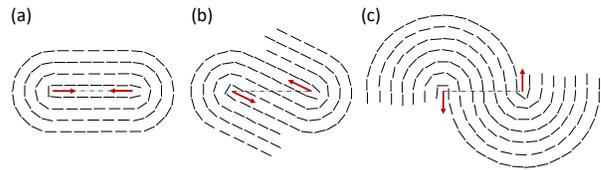}
    \caption{Examples of a pair of $+1/2$ defects with fixed Euclidean distance between the defect cores. The particle orientations are illustrated by the headless arrows. The defect orientation and position are represented by the red arrow and its tail. The number of layers $L$ between a pair of +1/2 defects from (a-c) are respectively $L=0, 3, 7$.}
    \label{fig:3}
\end{figure}

Since the head to tail alignment is the most stable configuration (Fig. \ref{fig:2}), string-like structures with alternating orientation at their side are expected to occur (see also Fig. \ref{fig:4}). Therefore, we evaluate the orientational order of the particles via the nematic order where the head and tail of the vector are treated similarly. The local order parameter tensor $\bm{Q}_i$ is calculated by averaging the orientation of the particles $i$ and its coordinated particles $N_i$: 
	\begin{equation}
	\bm{Q}_i=\frac{1}{1+N_i} \sum_{j=i, j \in N_i} (\bm{n}_j \bm{n}_j - \frac{1}{3}\bm{\delta})
	\end{equation}
The nematic order parameter $s$ is then three halves the positive eigenvalue of $\bm{Q}_i$. The position of the topological defect of the director field is approximated via data of  local nematic order parameter and winding number. The orientation of a defect is simply obtained by averaging the nematic orientation in the defect core region. In particular, for the comet-like $+1/2$ defect, the defect orientation is defined from the head to the tail of the comet. Figure \ref{fig:3} displays some configurations of a pair of $+1/2$ defects whose orientation is almost anti-parallel. Regarding the distance of the defects, aside from the ordinary Euclidean distance, we implement another parameter taking into account the orientation of the defects. The number of layers $L$ between a pair of $+1/2$ defects intrinsically includes the information of the relative position and orientation of a pair of defects, estimated by $L=\left \lfloor{ d\sin{\alpha}/h  }\right \rfloor $ where $d$ is the distance of the defect cores, $\alpha$ is angle between the defect direction and the distance vector of the defect cores, and $h$ is the thickness of two consecutive layers. It is shown in Fig. \ref{fig:3} that the number of layer between a pair of $+1/2$ defects varies although the Euclidean distance is fixed.

\section{\label{sec:Results}Results} 
\subsection{\label{sec:flat}Particles confined to planar geometry} 
\begin{figure}[ht]
    	\centering
   	\includegraphics[width=0.47\textwidth,trim=46mm 41mm 124mm 30mm ,clip]{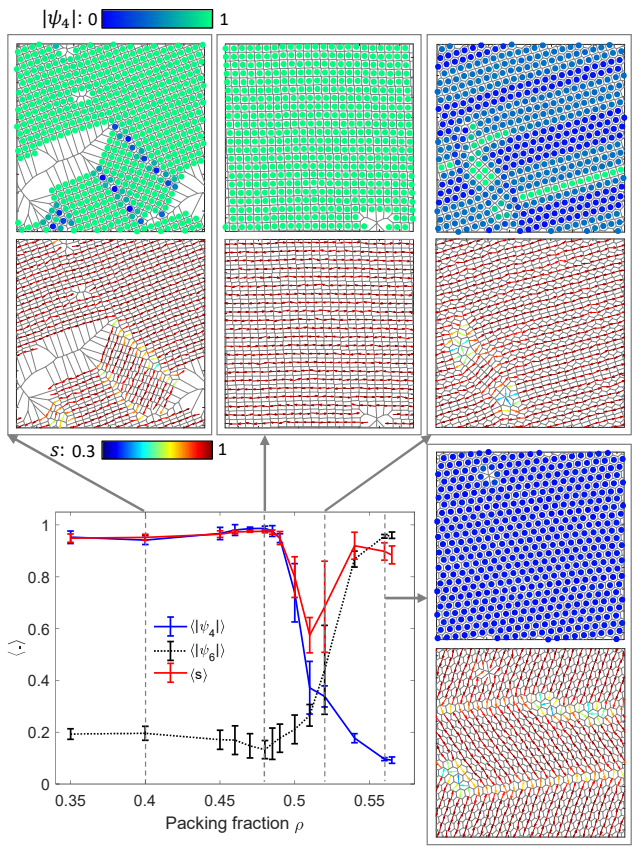}
    	\caption{The average properties $\langle |\psi_4|\rangle$, $\langle |\psi_6|\rangle$, and $\langle s\rangle$ of the dipole patchy particles in the planar geometry at various packing fractions. Each snapshot includes the particle positions with the local $|\psi_4|$ (filled circles), and particle orientations with the nematic order parameter $s$ (arrows) in the Voronoi cells. The length of the vector is taken as the particle diameter. The number of particles is $N=500$, at least 10 independent runs are conducted at each point.}
    	\label{fig:4}
	\end{figure}

The assembly of the dipole patchy particles on a flat space is investigated. Figure \ref{fig:4} shows the average properties in terms of bond-orientational order parameter and nematic order parameter, and the typical snapshots at various packing fractions $\rho$ from 0.35 to 0.565. The highly ordered, almost defect-free structures exhibited via the square lattice ($\langle |\psi_4|\rangle \approx 1$) and well aligned orientation ($\langle s\rangle \approx 1$) are more frequently obtained at the packing fraction comparable to $\rho \approx 0.48$. In these cases, the particle orientations are head to tail, whereas the side by side ones are anti-parallel. At $\rho<0.48$, grain boundaries emerge because an excess of voids induces more degrees of freedom of the grains. On the other hand, at dense packing $\rho>0.48$ clusters of 5- and/or 6-fold coordinated particles increase their sizes, thus the decrease in $\langle |\psi_4|\rangle$ and increase in $\langle |\psi_6|\rangle$ with $\rho$. We then perform simulations at the packing fraction $\rho=0.48$ so that a square lattice structure is formed. 

\subsection{\label{sec:S2}Particles confined to spherical surface} 
\begin{figure}[ht]
   \centering
   \includegraphics[width=0.46\textwidth,trim = 82mm 37mm 60mm 40mm, clip]{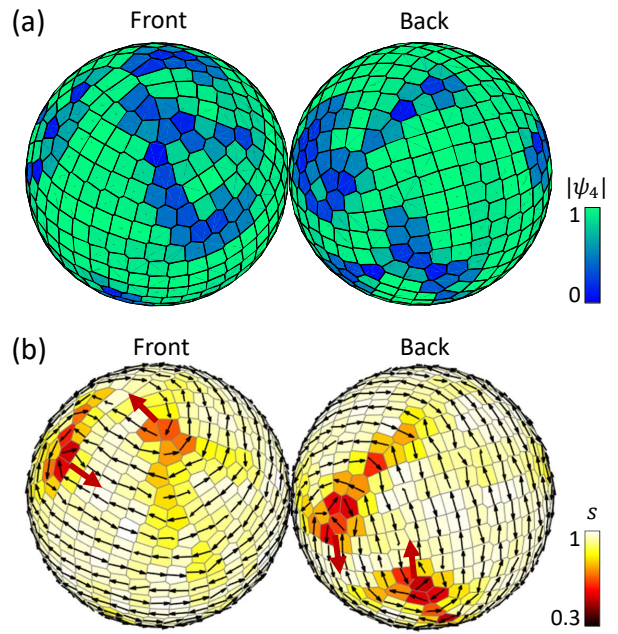}
   \caption{A self-assembly structure of N=500 dipole-like particles on a spherical surface with the local (a) 4-fold bond-orientational order parameter $|\psi_4|$ and (b) nematic order parameter $s$ inscribed in the Voronoi cells. The whole sphere consists of a front view and a back view of the two hemispheres connected by an imaginary hinge. The orientations of the particles are shown by the vectors of the same size with the particle's diameter. The positions of particles are at the midpoints of the vectors. The four big arrows illustrate the four $+1/2$ defects. The direction of the two $+1/2$ defects on the front (or back) are almost anti-parallel.}
   \label{fig:5}
\end{figure}

Simulations of dipole-like particles confined to a spherical surface are conducted at the conditions identical to the planar geometry case. Figure \ref{fig:5} illustrates the assembly of particles on the sphere at the volume packing for square lattice $\rho=0.48$. Different from the almost defect-free highly ordered structures in the flat space, the assembled structure on the sphere includes regions lacking both 4-fold order $|\psi_4|$ and nematic order $s$. There are four low nematic order regions whose winding number is $+1/2$, thus the Euler characteristic of $+2$ for the sphere is preserved. The anti-parallel direction of the defect core suggests that that there are two pairs of $+1/2$ defects. 

\subsubsection{Orientational order on spherical surface} 
\begin{figure} [ht]
   \centering
   \includegraphics[width=0.46\textwidth,trim =10mm 33mm 20mm 7mm, clip]{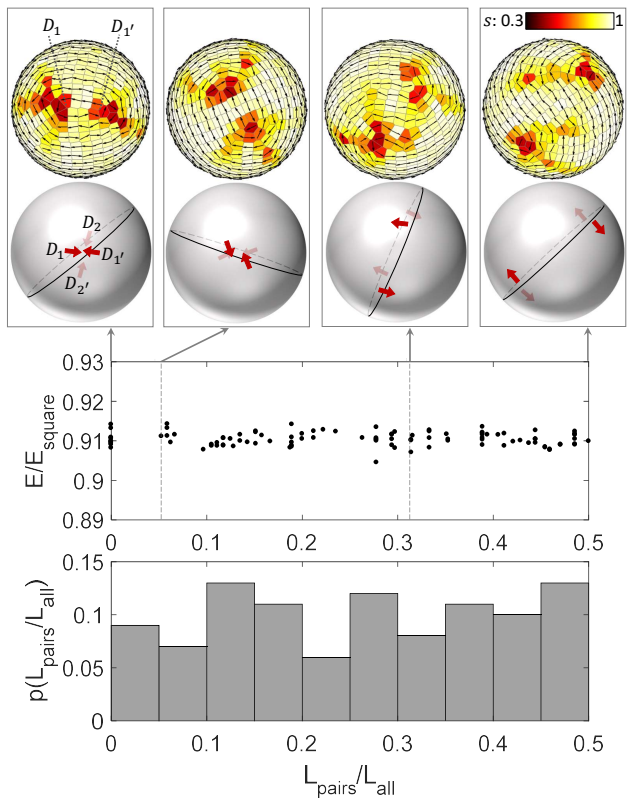}
   \caption{Distribution of the four half-strength defects for $N=500$. (Top) Some configurations of the assembled particles on the spherical surface. The hemisphere shows the particle orientation with local order parameter and a pair of defects. The locations of the pair $(D_1,D_{1'})$ in the front and pair $(D_2,D_{2'})$ in the back with the imaginary great circle is also given. From left to right the total number of layers of the pairs $(D_1,D_{1'})$, $(D_2,D_{2'})$ and the number of layers passing through the great circle are $(L_{pairs},L_{all})=(0,34), (2,34), (12,38), (20,40)$. (Bottom) The distribution of the relative position of the four defects shows similar probability. (Middle) The potential energy of spherical structures is almost independent from the defect's core position with coefficient of variation $0.2\%$. The energy is normalised with that of the perfect square lattice on planar geometry. The number of independent configurations is 100.}
   \label{fig:6}
\end{figure}

\begin{figure}[ht]
   \centering
   \includegraphics[width=0.46\textwidth,trim = 38mm 88mm 41mm 88mm, clip]{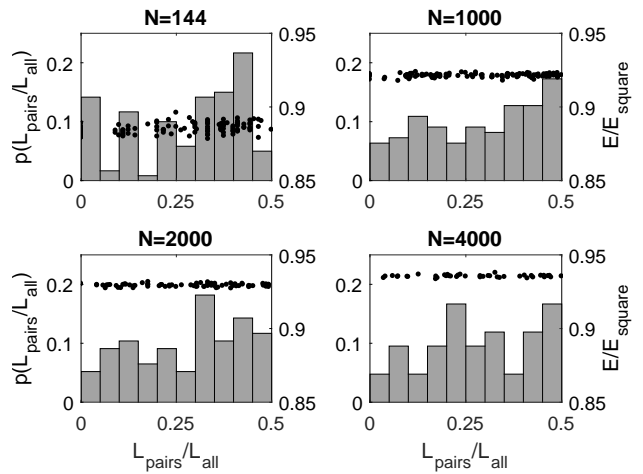}
   \caption{Distribution of the position of the four $+1/2$ defects (bar, to the left scale) and the potential energy (circle, to the right scale) at various system sizes. The number of independent runs for $N=144, 1000, 2000, 4000$ are $120, 110, 77, 42$, respectively.}
   \label{fig:7}
\end{figure}

Figure \ref{fig:6} depicts the positions and characteristics of the four $+1/2$ defects for the assembly of $N=500$ particles. At this size, the two pairs of defects $(D_1,D_{1'})$ and $(D_2,D_{2'})$ exhibit symmetric positions via the similar distance and the number of layers between each pair. A great circle can be drawn over the four $+1/2$ defect while minimising the deviation. The number of layers crossing through the whole great circle can be determined. The relative distance between the pairs of defects is evaluated as $L_{pairs}/L_{all}$. As shown in Fig. \ref{fig:6}, the distribution of the relative position of the defect cores has a similar probability. The potential energy of the system is almost identical regardless the defect cores are closely bound ($L_{pairs}/L_{all} \rightarrow 0$) or uniformly distributed on a great circle ($L_{pairs}/L_{all} \rightarrow 0.5$). Similar analysis for various system sizes is presented in Fig. \ref{fig:7}. For the small sphere size ($N=144$) there is some uneven distribution which is due to the discrete nature of the ratio $L_{pairs}/L_{all}$. As the sphere size increases, the ratio $L_{pairs}/L_{all}$ is less discrete, but defect is more likely to grow. Such kind of defect affects the position of the four $+1/2$, for example, the numbers of layers of the pair $(D_1,D_{1'})$ and $(D_2,D_{2'})$ are no longer equal. In general, the distribution of $L_{pairs}/L_{all}$ is not significantly different. The energy tends to be independent of the defect position, and approaches that of the square lattice in flat space. These findings suggest that the many states of the four $+1/2$ defects can be applied to a wide range of system size.  
 
\subsubsection{Positional order on spherical surface}
\begin{figure} [ht]
   \centering
   \includegraphics[width=0.46\textwidth, trim = 17mm 5mm 17mm 1mm, clip]{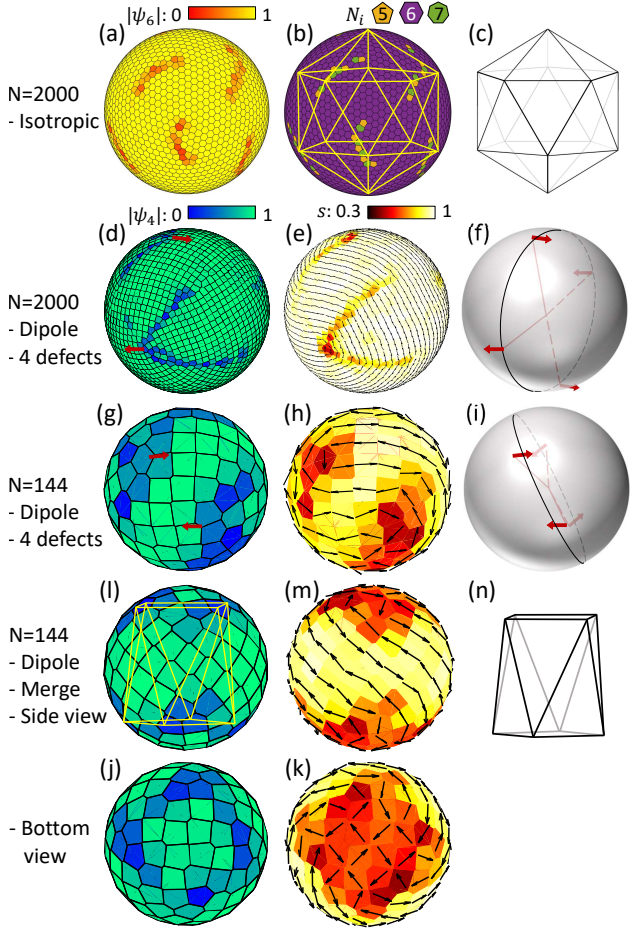}
   \caption{Arrangement of positional order of particles on a sphere. (a-c) A snapshot of $N=2000$ repulsive isotropic particles in terms of $|\psi_6|$, coordination number, and the icosahedral alignment of the defect scars which is the clusters of the 5- and 7-fold coordinated particles. (d-i) Snapshot of dipole patchy particles shown in $\psi_4$, $s$ and the position of the four $+1/2$ defects for $N=2000$ (d-f) and $N=144$ (g-i) particles; the red arrows mark the position and direction of the $+1/2$ defect. (l-k) The antiprism structure for $N=144$ particles in different view points}
   \label{fig:8}
\end{figure}

As mentioned elsewhere, perfect crystals confined to spherical surface do not exist; topological defects in the form of disclinations are required due to the topology of the sphere. We determine the disclinations by means of the local $|\psi_4|$. A particle of low $|\psi_4|$ indicates that either it has a non-square local structure or its coordination number is not equal to four. Therefore, the number of low $|\psi_4|$ particles is somewhat proportional to the number of disclinations as long as a suitable threshold of $|\psi_4|$ is chosen. This is similar to the correspondence between the clusters of low $|\psi_6|$ particles and the scars of 5-, 7-fold coordinated particles for the hexagonal lattice, as shown in Fig. \ref{fig:8}a-b. Regarding the square lattice on large sphere size ($N=2000$, Fig. \ref{fig:8}d-f), the one-dimensional low $|\psi_4|$ regions are observed as grain boundary scars. There are two grain boundary scars emerging from each $+1/2$ defects, which makes eight scars in total, then disappearing within the otherwise square lattice particles. The shape of the two  scars for each $+1/2$ defect is always like lines connected by a certain angle. In contrast, the twelve scars of isotropic particles have more degrees of freedom on the arrangement the scars (Fig. \ref{fig:8}a-c) \cite{bowick_two-dimensional_2009}. For the small sphere ($N=144$) two cases may occur. When the $+1/2$ defects are clearly observed as presented in Fig. \ref{fig:8}g-i, the two scars around the $+1/2$ defect reduces its size, and become two points if only the particles with $|\psi_4|<0.2$ are considered. In the case when the $+1/2$ defects are ill-defined, the particle orientations are well aligned around the sphere's equator while lacks of order at the two poles (Fig. \ref{fig:8}l-k). At each pole, the four lowest $|\psi_4|$ particles create a square. The eight lowest $|\psi_4|$ points form a square antiprism. This structure is somewhat in agreement with the conjecture of minimum disclinations for square lattice by using Euler theory, that at least eight 3-fold coordinated particles is required on a sphere (see Appendix \ref{sec:Appendix_Euler}). 

\begin{figure} [ht]
   \centering
   \includegraphics[width=0.45\textwidth, trim = 30mm 84mm 30mm 85mm, clip]{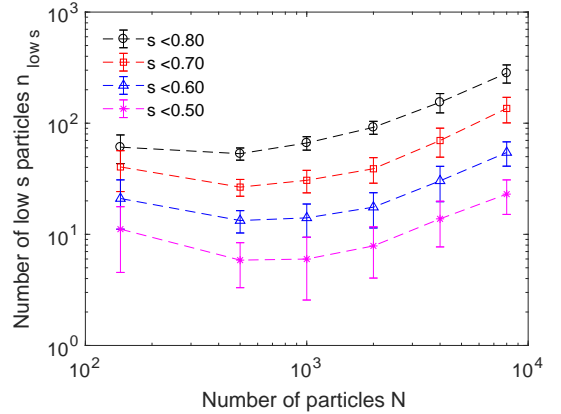}
   \caption{The relation of the number of low nematic order particles to the total number of particles $N$ at different constraints of $s$ value.}
   \label{fig:10}
\end{figure}

Figure \ref{fig:9}-\ref{fig:10} show the dependence of the number of disordered particles on the system size. The relation of the number low $|\psi_4|$ to the number of particles is displayed in Fig. \ref{fig:9}. Although the threshold of $|\psi_4|$ varies, the number of non-square particle $n_{\mathrm{low } |\psi_4|}\propto N^{0.5}$, implying that the grain boundary scars increase almost linearly with the radius of the sphere. Such linear dependence has been observed for isotropic particles on a sphere, that the excess disclinations including 5- and 7-fold particles linearly increases with the sphere radius \cite{bowick_interacting_2000,bausch_grain_2003}. The dependence of the number of low nematic order particles on system size is different from that of $|\psi_4|$ in terms of scale and trend (Fig. \ref{fig:10}). The number of low $s$ particles increases with the system size for the threshold $s<0.8$ and $s<0.7$. However, as the threshold is lowered to $s<0.6$, the number of disorder particles below the threshold reduced to a few dozens regardless the size of the sphere. As given in Fig. \ref{fig:8}(d-e), there is a strong connection between the low $|\psi_4|$ and low $s$ regions. However, the orientations of the particles at the grain boundary scars are distorted at a different level: the nematic order near the $+1/2$ defect core is significantly lower than the others. In other words, the distinguishing behaviour of the number of non-square particles and the number of disorder nematics to the system size can be originated from the characteristic of the 1-dimensional grain boundary scar and the 0-dimensional $+1/2$ disclination, plus the interaction between them.     

\begin{figure} [ht]
   \centering
    \includegraphics[width=0.45\textwidth, trim = 30mm 84mm 30mm 85mm, clip]{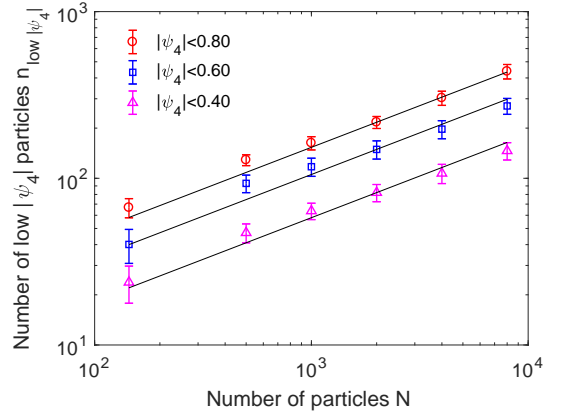}
    \caption{The relation of the number of low 4-fold order particles to the total number of particles $N$ at different constraints of $|\psi_4|$ value. The lines are guidance for eye and has slope coefficient of 0.5.}
   \label{fig:9}
\end{figure}

\section{\label{sec:Summary}Discussion and Summary}
We have performed the assembly of dipole patchy particles on a spherical surface. Defects appear as a requirement by the topology of the sphere. As for the orientations of the particles, there are many states of the four $+1/2$ defects near a great circle. The appearance of many stable states of four half-strength defects of director field is also observed in spherical smectics \cite{xing_topology_2009,serafin_topology_2018}, spherical nematics in the extreme elastic constant limit \cite{shin_topological_2008}, block copolymer assembly \cite{chantawansri_self-consistent_2007}. In those studies, the four $+1/2$ defects can be explained by the cut-and-rotate on a sphere of two $+1$ defects. This operation seems to applied well to our system. However, we have not found the clear appearance of a two $+1$ defects in the form of concentric layers of particle orientation; and a deformed form is found instead (see Fig. \ref{fig:3}a,\ref{fig:6}). Such difference is possibly brought about by either the non-zero temperature simulation or the soft, anisotropic interacting potential used in our study. Regarding the topological defect of the square lattice order, the emergence of the grain boundary scars and its linear relation with the sphere radius are moderately analogical to that of the hexagonal lattice \cite{bowick_two-dimensional_2009}. The grain boundary scars are revealed to be associated with the four $+1/2$ defects. It is also possible that such mutual interplay between the positional order and the orientational order suppresses the defect-defect interaction, leading to the many position of the four $+1/2$ defects. 

Although the 4-fold bond-orientational order parameter $\psi_4$ is sufficient enough to evaluate the disclination, it is still inadequate to determine the exact 3-fold or 5-fold coordinated particles in a square lattice. One should consider further development for the quadrilateral mesh, especially the irregular regions near a defect. This also gives rise to a question on the defect analysis for more complex structures. 

The complex interaction between the positional order and orientational order of particles on a sphere is of fundamental interest for two-dimensional melting. Tuning the potential, \textit{e.g.} varying the current short-range interaction to the long-range one, or the type of patchy particle, deformability of the surface, may bring better understanding on the role of the energy-driven factor on the in-plane order. On the materials science aspect, knowing the precise defect structure is important for fabricating building block, for example, the eight scars in the study can be functionalised by chemical linkers, then served as interacting sites. 

\section{\label{sec:Appendix}Appendix} 
\subsection{\label{sec:Appendix_Euler}Derivation of minimum number of positional defects from Euler theorem}
Suppose a closed surface is facetted and divided into a number of $V$ vertices, $E$ edges, $F$ faces. Euler theorem states that
	\begin{equation}
	V-E+F=\chi
	\end{equation}
where the Euler characteristic $\chi=2(1-g)$, and $g$ is the genus or the number of holes of a closed surface, for example $\chi=2$ for all polyhedra because they are topologically equal to a sphere \cite{hyde_language_1997}. One may image that such facetted surface is similar to a net embedded on the surface. If we restrict the ring of the net has $c$ vertices, then  
	\begin{equation}
	E=cV/2
	\end{equation}
On the other hand, each node or vertex of the net may have $z$ connection with the others \textit{i.e.} there would be $N_z$ dual polygons with $z$ sides. One may find
 	\begin{align}
	&F=\sum_{z}{N_z} \\
	&\sum_{z}{zN_z} =cV
	\end{align}
After substituting we obtain
	\begin{equation}
	\sum_{z}{ \left( \frac{2c}{c-2} -z \right)N_z } =\frac{2c}{c-2} \chi 
	\end{equation}
For example, applying the above equation for the triangular lattice ($c=3$) and quadrilateral lattice ($c=4$) gives $\sum_{z}{ \left( 6-z \right)N_z } =6\chi$ and $\sum_{z}{ \left( 4-z \right)N_z } =4\chi$, respectively. It means on a sphere, the minimum number of disclinations is twelve 5-fold coordinated nodes for triangular lattice and eight 3-fold coordinated nodes for quadrilateral one.  

\section*{Conflicts of interest}
There are no conflicts to declare.





\bibliography{MyLibrary} 
\bibliographystyle{unsrt}

\end{document}